\documentclass[12pt,preprint]{aastex}
\input psfig.tex

\def\eg{{\it e.g.}\ }

\def\ie{{\it i.e.}\ }
\newcommand{\be}{\begin{equation}}
\newcommand{\ee}{\end{equation}}
\def \pmbmath{\mathpalette\pmbmathaux}
\def \pmbmathaux#1#2{
         \pmbtext{$#1#2$}}
\def \pmbtext#1{\leavevmode
     \setbox0\hbox{#1}
     \kern-0,2pt \copy0 \kern-\wd0
     \kern0,4pt \copy0 \kern-\wd0
     \kern-0,2pt \raise0,3pt \box0}
\def\kpa{{k_\parallel}}
\def\kpe{{k_\perp}}
\slugcomment{\today}

\shorttitle{Diffusion equations for MHD turbulence}
\shortauthors{Galtier}

\begin{document}

\title{Nonlinear diffusion equations for anisotropic MHD turbulence with cross-helicity}

\author{S\'ebastien Galtier\altaffilmark{1,2} and \'Eric Buchlin\altaffilmark{1}}
\affil{Univ Paris-Sud, Institut d'Astrophysique Spatiale, UMR8617, b\^at. 121, F-91405 Orsay, France}
\email{sebastien.galtier@ias.fr}

\altaffiltext{1}{CNRS, Orsay, F-91405}
\altaffiltext{2}{also at Institut universitaire de France}

\begin{abstract}
Nonlinear diffusion equations of spectral transfer are systematically derived for anisotropic magnetohydrodynamics 
in the regime of wave turbulence. The background of the analysis is the asymptotic Alfv\'en wave turbulence  
equations from which a differential limit is taken. The result is a universal diffusion-type equation in 
${\bf k}$-space which describes in a simple way and without free parameter the energy transport 
perpendicular to the external magnetic field ${\bf B_0}$ for transverse and parallel fluctuations. These 
equations are compatible with both the thermodynamic equilibrium and the finite flux spectra derived by 
Galtier et al. (2000); it improves therefore the model built heuristically by Litwick \& Goldreich (2003) for 
which only the second solution was recovered. This new system offers a powerful description of a wide 
class of astrophysical plasmas with non-zero cross-helicity. 
\end{abstract}


\section{Introduction}

The observations of astrophysical plasmas by various spacecrafts have added substantially to our 
knowledge of magnetohydrodynamic (MHD) turbulence. Among the different media widely analyzed 
like the interstellar medium \citep{elmegreen}  or the Sun's atmosphere \citep{chae}, the solar wind is 
certainly the most interesting plasma since direct measurements are possible. This unique situation 
in astrophysics allows us to probe deeply the nature of the fluctuations and to investigate for example 
the origin of anisotropy \citep{Bill96,Alexakis,podesta}, to evaluate the mean energy dissipation 
rate \citep{macbride}, to detect intermittency \citep{salem}, or to analyze the transition to the regime 
of dispersive turbulence characterized by a steepening of the magnetic field fluctuations spectrum with 
a power law index going from $-5/3$, at frequencies lower than $1$Hz, to indices lying around $-2.5$ at 
higher frequencies \citep{Galtier06a,Smith06,Galtier2008,Sahraoui}. 

The low solar corona provides a second interesting example where it is believed that MHD turbulence 
plays a central role in the dynamics and the small scale heating. For example, in active region loops 
spectrometer analyses revealed non-thermal velocities reaching sometimes $50$ km/s \citep{chae}; 
this line broadening is generally interpreted as unresolved turbulent motions with length scales smaller 
than the diameter of coronal loops which is about one arcsec and timescales shorter than the exposure time 
of the order of few seconds. Turbulence is evoked in the solar coronal heating problem since it offers a 
natural process to produce small scale heating \citep{HP92,Galtier1999,Cranmer10}. 
Weak MHD turbulence is now proposed has a possible regime for some coronal loops since a very small 
ratio is expected between the fluctuating magnetic field and the axial component \citep{Rappazzo}. Inspired 
by the observations and by recent direct numerical simulations of three-dimensional MHD turbulence 
\citep{Bigot08a}, 
an analytical model of coronal structures has been proposed \citep{Bigot08b} where the heating is seen as 
the end product of a wave turbulent cascade. Surprisingly, the heating rate found is non negligible and may 
explain the observational predictions. 

A third example where MHD turbulence seems to be fundamental is given by the upper solar corona which 
makes a connection between the lower corona and the stationary solar wind. Observations reveal that the 
heating in this region affects preferentially the ions in the direction perpendicular to the mean magnetic field. 
The electrons are much cooler than the ions, with temperatures generally less than or close to $10^6$K 
\citep{david}. Additionally, the heavy ions become hotter than the protons within a solar radius of the coronal 
base. Ion cyclotron waves could be the agent which heats the coronal ions and accelerates the fast wind. 
Naturally the question of the origin of these high frequency waves arises. Among different scenarios, 
turbulence appears to be a natural and efficient mechanism to produce ion cyclotron waves. In this case, the 
Alfv\'en waves launched at low altitude with frequencies in the MHD range, would develop a turbulent 
cascade to finally degenerate and produce ion cyclotron waves at much higher frequencies. In that context, 
the wave turbulence regime was considered in the weakly compressible MHD case at low-$\beta$ plasmas 
(where $\beta$ is the ratio between the thermal and magnetic pressure) in order to analyze the nonlinear 
three-wave interaction transfer to high frequency waves \citep{chandran}. The wave turbulence calculation 
shows -- in absence of slow magnetosonic waves -- that MHD turbulence is a promising explanation for 
the anisotropic ion heating. 

MHD turbulence modeling is the main tool to investigate the situations previously discussed. Although 
it cannot be denied that numerical resources have been significantly improved during the last decades
\citep{mininni}, direct numerical simulations of MHD equations are still limited for describing highly turbulent 
media. For that reason, shell cascade models are currently often used to investigate the small scale coronal 
heating \citep{buchlin} and its impact in terms of spectroscopic emission lines. Transport equations are also 
used for example in the context of solar wind acceleration in the extended solar corona \citep{Cranmer03}. 
The {\it ad hoc} model is an advection-diffusion equation for the evolution of the energy spectrum whose 
inspiration is found in the original paper by \cite{leith}. It is also a cascade model where the locality of the 
nonlinear interactions is assumed but where the dynamics is given by a second-order nonlinear partial 
differential equation whereas we have ordinary differential equations for shell models. 

In next Section, the origin of the Leith's model is discussed and in Section \ref{AWT} the Alfv\'en wave 
turbulence equations are reminded in the case of non-zero cross-helicity. In Section \ref{DA}, the differential 
limit is taken on the previous wave turbulence equations and the associated nonlinear diffusion equations 
for anisotropic MHD turbulence are systematically derived. Finally, a conclusion is developed in the last Section.

\section{Leith's model}

A theoretical understanding of the statistics of turbulence and the origin of the power law energy spectrum, 
generally postulated from dimensional considerations {\it \`a la} Kolmogorov, remains one of the outstanding 
problems in classical physics which continues to resist modern efforts at solution. The difficulty lies in the 
strong nonlinearity of the governing equations which leads to an unclosed hierarchy of equations. Faced 
with that situation different models have been developed like closure models in Fourier space for 
hydrodynamic and magnetohydrodynamic turbulence \citep{kraichnan63,orszag}. In the meantime --
and following an approach often fruitful in radiation and neutron transport theory \citep{davidson} -- 
Leith introduced the idea of a diffusion approximation to inertial energy transfer in isotropic turbulence
\citep{leith}. This new class of {\it ad-hoc} models describes the time evolution of the spectral energy density, 
$e({\bf k})$, for originally an isotropic three-dimensional incompressible hydrodynamic turbulence, in terms 
of a partial differential equation by making a diffusion approximation to the energy transport process in the 
${\bf k}$--space representation. Ignoring forcing and dissipation the three-dimensional isotropic Navier-Stokes 
equations read in Fourier space
\be
{\partial e({\bf k}) \over \partial t} =- \nabla \cdot {\bf {\cal F}} 
= - {1 \over k^2} {\partial \over \partial k} \left( k^2 {\cal F}_r \right) \, .
\label{eq1}
\ee
The radial component of the energy flux vector is modeled as 
\be
{\cal F}_r = - D(k) {\partial e({\bf k}) \over \partial k} \, , 
\ee
where $D(k)$ is a diffusion coefficient that remains to be determined. It is straightforward to show  
dimensionally that the diffusion coefficient scales as 
\be
D \sim {k^2 \over \tau} \, , 
\label{diff}
\ee
where $\tau$ is the typical transfer time of the Navier-Stokes equations which can be identified as the eddy 
turnover time $\tau_{eddy}$. Therefore, we may evaluate this time as
\be 
\tau = \tau_{eddy} \sim {1 \over k \sqrt{ek^3}} \, .
\label{time}
\ee
We remind that the total kinetic energy per mass is by definition $\int e({\bf k}) d {\bf k} = \int E(k) dk$.
After substitution of (\ref{time}) into (\ref{diff}) it is possible to rewrite (up to a factor) the model equation 
(\ref{eq1}) for the omnidirectional spectrum \citep{leith}
\be
{\partial E(k) \over \partial t} = {\partial \over \partial k} \left( k^{11/2} E^{1/2} 
{\partial \over \partial k} \left( E/k^2 \right) \right) \, , 
\label{eq2}
\ee
which is commonly named the Leith's equation.
Beyond its relative simplicity, equation (\ref{eq2}) exhibits several important properties like the preservation 
after time integration of a non negative spectral energy and the production of the Kolmogorov spectrum in the 
inertial range which corresponds to a finite energy flux solution. It is straightforward to prove that by imposing 
a constant energy flux in the inertial range, namely 
\be 
k^{11/2} E^{1/2} {\partial \over \partial k} \left( E/k^2 \right) = {\rm constant} \, . 
\ee
If we look for power law solutions, $E \sim k^x$, then the unique solution that emerges is $x=-5/3$. 
Note that this equation may also exhibit an anomalous scaling during the non stationary phase with 
a steeper power law \citep{colm}. 

A generalization of the Leith's model to three-dimensional isotropic MHD turbulence was proposed by 
\citet{zhou}. (Note that \citet{Iro} proposed the first such a model for MHD from which the $-3/2$ spectrum 
was derived). The main modification happens in the evaluation of the transfer time $\tau$ for which a 
combination of the eddy turnover time $\tau_{eddy}$ and the Alfv\'en time $\tau_A$ is proposed. The 
phenomenological evaluation of the transfer time allows the recovering of either the Heisenberg--Kolmogorov 
($-5/3$) or the Iroshnikov--Kraichnan ($-3/2$) spectrum when the ratio $\tau_{eddy} / \tau_A$ is respectively 
much less or much larger than one \citep{K41,Hen,Iro,Krai}. 
The model was also adapted to the case of a non-zero cross-helicity for which a distinction 
was made between the Els\"asser energies $E^+$ and $E^-$. The generalization of the Leith's model to 
the more realistic situation of anisotropic MHD turbulence where an external magnetic field ${\bf B_0}$ is 
imposed was proposed only recently \citep{Bill09}. 
As already announced by \citet{zhou} the departure from the assumption of isotropic turbulence generates
a difficult mathematical treatment since, in particular, a diffusion tensor is expected instead of a scalar. 
Another difficulty comes from the locality of the nonlinear interaction which is assumed in the isotropic case:
when a mean magnetic field is imposed the situation is different since a reduction of nonlinear transfers occurs 
along ${\bf B_0}$. In terms of triads, ${\bf k} = \pmbmath{\kappa} + {\bf L}$, it means that one of the wavevectors, 
say $\pmbmath{\kappa}$, is mainly oriented transverse to ${\bf B_0}$. The sophisticated model proposed 
by \cite{zhou} is an attempt to describe such a nontrivial dynamics. 

The case of Alfv\'en wave turbulence for which a relatively strong $B_0$ is required is an important limit for 
which a rigorous analysis is possible \citep{Galtier2000}. The wave kinetic equations derived are a set of 
coupled integro-differential equations which are not obvious to simulate numerically in the most general 
case \citep{Galtier2000,Bigot08c}. This 
remark was a motivation for deriving a model made of two coupled diffusion equations which describe 
Alfv\'en wave turbulence with a non-zero cross-helicity \citep{LG03}. These model equations are able to 
recover the finite flux spectra which are exact solutions of the wave kinetic equations \citep{Galtier2000}. 
In the present paper, it is shown that a set of two coupled nonlinear diffusion equations may be derived 
{\it systematically} from the asymptotic equations of Alfv\'en wave turbulence by taking a differential limit. An 
important difference is found between the nonlinear diffusion equations derived here and the model proposed 
by \cite{LG03}. The main physical problem is the inability for the model to reproduce the thermodynamic 
equilibrium solutions which are exact solutions of the wave turbulence equations. It is believed that the 
higher degree of accuracy of the new system offers a powerful description of a wide class of astrophysical 
plasmas with non-zero cross-helicity.

\section{Asymptotic theory of Alfv\'en wave turbulence}
\label{AWT}

The wave turbulence theory for three-dimensional incompressible MHD was derived rigorously by 
\cite{Galtier2000}. It is a perturbative theory which necessitates heavy calculations which will not be 
reproduced here. We refer the reader to the original paper for a global explanation or to two satellite 
papers where simplified approaches are adopted \citep{Galtier2002,Galtier06b}. Since it is important 
to understand the wave turbulence equations from which our analysis will start, we shall remind below 
the main steps in their derivation. 

The inviscid incompressible three-dimensional MHD equations read 
\begin{eqnarray}
\partial_t {\bf z}^s - s {\bf B_0} \cdot \mathbf{\nabla} {\bf z}^s 
&=& - {\bf z}^{-s} \cdot \mathbf{\nabla}  {\bf z}^s - \mathbf{\nabla} P_* \, , \\
\mathbf{\nabla} \cdot {\bf z}^s &=& 0 \, ,
\label{ZMHDeq}
\end{eqnarray}
where ${\bf z}^s = {\bf v} + s {\bf b}$ are the Els\"asser fields ($s=\pm$), ${\bf v}$ is the fluid velocity, 
${\bf b}$ is the magnetic field in velocity units, ${\bf B_0}$ is a uniform magnetic field (in velocity units, 
\ie the Alfv\'en speed) and $P_*$ is the total (thermal plus magnetic) pressure. We assume that the 
uniform magnetic field is relatively strong ($B_0 \gg z^s$) and that MHD turbulence is dominated by a 
wave dynamics for which the nonlinearities are weak. In such a limit, a small parameter $\epsilon$ 
may be introduced formally to measure the strength of the nonlinearities. Then, we obtain for the 
jth-component
\begin{eqnarray}
\left(\partial_t -s B_0 \partial_\parallel \right)z_j^s 
= -\epsilon z^{-s}_m \partial_m z^s_j - \partial_j P_* \  ,
\label{etf}
\end{eqnarray}
where the Einstein's notation is used for the indices. Note that the parallel direction ($\parallel$) corresponds 
to the direction along ${\bf B_0}$. We shall Fourier transform such equations with the following definition for 
the Fourier transform of the Els\"asser field components $z^s_j(\mathbf{x},t)$:
\begin{eqnarray}
z^s_j({\bf x},t) = \int a^s_j(\mathbf{k},t) \, e^{i(\mathbf{k} \cdot {\bf x}+s \omega_k t)} \, d\mathbf{k} \, ,
\end{eqnarray}
where $\omega_k=B_0 \kpa$ is the Alfv\'en frequency. The quantity $a^s_j(\mathbf{k},t)$ is the wave 
amplitude in the interaction representation, hence the factor $e^{is\omega_k t}$. Then, the Fourier 
transform of equation (\ref{etf}) gives 
\begin{eqnarray}
\partial_t a^s_j(\mathbf{k})=-i\epsilon k_m P_{jn} \int\int a^{-s}_m(\pmbmath{\kappa}) a^s_n(\mathbf{L}) 
e^{is\Delta\omega t} \delta_{\mathbf{k},\pmbmath{\kappa}\mathbf{L}}
d\pmbmath{\kappa} \, d\mathbf{L} \ .
\label{amplitude}
\end{eqnarray}
Here, $P_{jn}$ is the projector on solenoidal vectors such that $P_{jn}(\textbf{k})=\delta_{jn}-k_j k_n/k^2$; 
$\delta_{\mathbf{k},\mathbf{\kappa}\mathbf{L}}=\delta(\mathbf{k}-\pmbmath{\kappa}-\mathbf{L})$ reflects 
the triadic interaction, and $\Delta\omega=\omega_L-\omega_k-\omega_\kappa$ is the frequency mixing.
The appearance of an integration over wave vectors $\pmbmath{\kappa}$ and $\mathbf{L}$ is directly linked 
to the quadratic nonlinearity of equation (\ref{etf}) (as a result of Fourier transform of a correlation product).

\begin{figure}
\plotone{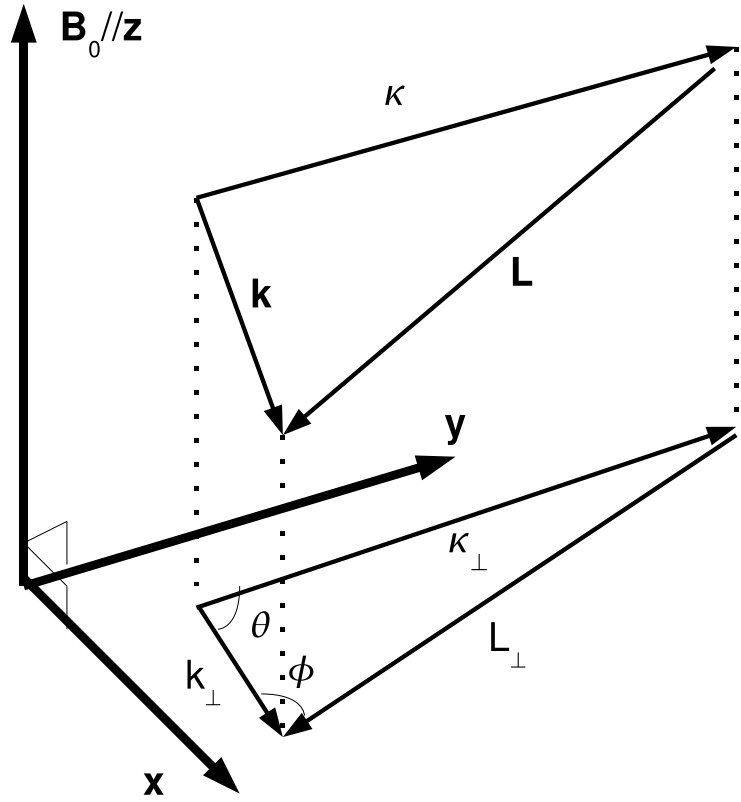}
\caption{Triadic interaction $\mathbf{k}=\pmbmath{\kappa} + \mathbf{L}$ and its projection in the plane 
perpendicular to $\mathbf{B}_0$.}
\label{Fig1}
\end{figure}

Equation (\ref{amplitude}) is nothing else than the compact expression of the incompressible MHD equations 
when a strong uniform magnetic is present. It is the point of departure of the wave turbulence formalism which 
consists in writing equations for the long time behavior of second order moments. In such a statistical 
development, the time-scale separation, $\tau_{A}/\tau_{eddy} \ll 1$ (with $\tau_A = 1/\omega_k$ and 
$\tau_{eddy} = 1/\kpe z^s$), leads asymptotically to the destruction of some nonlinear terms -- including the 
fourth order cumulants -- and only resonance terms survive \citep{Galtier2000,Galtier2009a}. It leads to 
a natural asymptotic closure for the moment equations. In such a statistical development, the following general 
definition for the total (shear-- plus pseudo--Alfv\'en wave) energy spectrum is used;
\begin{equation}
\langle a^s_j(\mathbf{k}) a^s_j(\mathbf{k^{\prime}}) \rangle = 
E^s(\mathbf{k}) \, \delta(\mathbf{k}+\mathbf{k^{\prime}}) / k_\perp \, ,
\label{esh0}
\end{equation}
where $\langle \rangle$ stands for an ensemble average and $k_\perp=\vert \mathbf{k_\perp} \vert$.
In absence of magnetic helicity and in the case of an axially symmetric turbulence, the asymptotic equations 
simplify. For shear-Alfv\'en waves\footnote{We recall that shear-Alfv\'en and pseudo-Alfv\'en waves are the 
two kinds of linear perturbations about the equilibrium, the latter being the incompressible limit of slow 
magnetosonic waves.}, the energy spectrum is given by  
\be
E_{shear}^s(k_\perp,k_\parallel)=g(k_\parallel) E^s_\perp(k_\perp) \, ,
\label{esh}
\ee
where $g(k_\parallel)$ is a function fixed by the initial conditions ({\it ie.} there is no energy transfer along 
the parallel direction). In the limit $k_\perp \gg k_\parallel$, the transverse part obeys the following nonlinear 
equation (the small parameter $\epsilon$ is now included in the time variable and the limits of the integration
are explicitly written)
\begin{eqnarray}
\partial_t E^s_\perp(k_\perp)=
{\pi \over B_0} \int_0^{+\infty}  \int_0^{+\infty}  \cos^2\phi \sin\theta  \, \frac{k_\perp}{\kappa_\perp}
E^{-s}_\perp(\kappa_\perp) \left[k_\perp E^s_\perp(L_\perp)-L_\perp E^s_\perp(k_\perp)\right]
d\kappa_\perp dL_\perp , 
\label{shearEq}
\end{eqnarray}
where $\phi$ is the angle between $\mathbf{\kpe}$ and $\mathbf{L_\perp}$, and $\theta$ is the angle 
between $\mathbf{k_\perp}$ and $\pmbmath{\kappa}_\perp$ with the perpendicular wave vectors 
satisfying the triangular relation $\mathbf{k_\perp} = \mathbf{L_\perp} + \pmbmath{\kappa}_\perp$ 
(see Figure \ref{Fig1}). 
Note that from the axisymmetric assumption, the azimuthal angle integration has already
been performed, and we are only left with an integration over the absolute values of the
two wave numbers, $\kappa_\perp = |\pmbmath{\kappa}_\perp|$ and $L_\perp = |\mathbf{L_\perp}|$.
In the same way, equations can be written for pseudo-Alfv\'en waves which are 
passively advected by shear-Alfv\'en waves, namely 
\begin{eqnarray}
\partial_t E^s_\parallel(k_\perp) = {\pi \over B_0} \int_0^{+\infty} \int_0^{+\infty} 
\sin\theta \, \frac{k_\perp}{\kappa_\perp} E^{-s}_\perp(\kappa_\perp) 
\left[k_\perp E^s_\parallel(L_\perp) -L_\perp E^s_\parallel(k_\perp)\right] d\kappa_\perp dL_\perp ,
\label{pseuEq}
\end{eqnarray}
with by definition
\be
E_{pseudo}^s(k_\perp,k_\parallel)={\tilde g}(k_\parallel) E^s_\parallel(k_\perp) \, ,
\ee
where ${\tilde g}(k_\parallel)$ is a function determined by the initial condition. 

Equations (\ref{shearEq}) and (\ref{pseuEq}) are asymptotically exact. These master equations of Alfv\'en 
wave turbulence describe the nonlinear evolution of MHD turbulence in the presence of a strong uniform 
magnetic field, with non-zero cross-helicity and zero magnetic helicity\footnote{We refer to the original paper
\citep{Galtier2000} for a discussion about the domain of applicability in terms of wavevectors of the Alfv\'en 
wave turbulence regime.}. In the limit $k_\perp \gg k_\parallel$ considered, equations (\ref{shearEq}) and 
(\ref{pseuEq}) describe respectively the dynamical evolution of transverse and parallel fluctuations.

\section{Differential limit for strongly local interactions}
\label{DA}

We shall take a differential limit of equations (\ref{shearEq}) and (\ref{pseuEq}) for strongly local interactions 
\citep{Dyachenko}. It is important to note that by taking this limit we shall extract a subset of the full set of 
interactions that is present in the Alfv\'en wave turbulence equations (\ref{shearEq})--(\ref{pseuEq}). In terms 
of triads, strongly local interactions\footnote{Strongly {\it nonlocal} interactions is another interesting limit from
which we may derive turbulent viscosities in the wave turbulence regime \citep{Bigot08b} or in the strong 
turbulence regime \citep{pouquet}. In the latter case, an EDQNM closure model was used and the main 
application was the dynamo problem whereas in the former case the application was the solar corona with 
coronal loops and coronal holes.} means that we will only retain triangles which are approximately equilateral. 
(Note that the locality concerns only perpendicular wavevectors.) Therefore, the differential limit will lead to an 
approximate description of Alfv\'en wave turbulence which is believed, however, sufficiently rich to reproduce 
the most important properties of the original system. 
The rigorous derivation will be presented only for shear-Alfv\'en waves since the generalization to pseudo-Alfv\'en 
waves is straightforward. Multiplying equation (\ref{shearEq}) by an unknown function $f(k_\perp)$ we obtain, 
after integration in $k_\perp$, 
\begin{eqnarray}
\partial_t \int_0^{+\infty} E^s_\perp(k_\perp) f(k_\perp) dk_\perp = 
\int_0^{+\infty} \int_0^{+\infty} \int_0^{+\infty} T(k_\perp,L_\perp,\kappa_\perp) f(k_\perp) 
dk_\perp d\kappa_\perp dL_\perp \, ,
\label{E1}
\end{eqnarray}
where
\be
T(k_\perp,L_\perp,\kappa_\perp) = {\pi \over B_0} \cos^2\phi \sin\phi  \, \frac{k_\perp L_\perp}{\kappa_\perp^2}
E^{-s}_\perp(\kappa_\perp) \left[k_\perp E^s_\perp(L_\perp)-L_\perp E^s_\perp(k_\perp)\right] \, .
\ee
Note the use of the triangle relation (see Figure \ref{Fig1}) 
\be 
 \sin\theta =  \sin\phi {L_\perp \over \kappa_\perp} \, .
\ee
It is convenient to introduce the following definition $S = \pi \cos^2\phi \sin\phi$. 
Then, by changing the name of indices one can write 
\be
\partial_t \int_0^{+\infty} E^s_\perp(k_\perp) f(k_\perp) dk_\perp = 
\label{E2}
\ee
$${1 \over 2} \int_0^{+\infty} \int_0^{+\infty} \int_0^{+\infty} [T(k_\perp,L_\perp,\kappa_\perp) f(k_\perp) + 
T(L_\perp,k_\perp,\kappa_\perp) f(L_\perp) ] dk_\perp d\kappa_\perp dL_\perp \, .$$
However, by symmetry we also have 
\be 
T(L_\perp,k_\perp,\kappa_\perp) = - T(k_\perp,L_\perp,\kappa_\perp) \, ,
\ee
which gives
\begin{eqnarray}
\partial_t \int_0^{+\infty} E^s_\perp(k_\perp) f(k_\perp) dk_\perp = {1 \over 2} 
\int_0^{+\infty} \int_0^{+\infty} \int_0^{+\infty} T(k_\perp,L_\perp,\kappa_\perp) (f(k_\perp) - f(L_\perp) ) 
dk_\perp d\kappa_\perp dL_\perp \, .
\label{E3}
\end{eqnarray}
For strongly local interactions, we have the following relations
\be
\kappa_\perp =  k_\perp (1 + \epsilon_\kappa) \, , \quad \vert \epsilon_\kappa \vert \ll 1 \, , 
\ee
\be
L_\perp = k_\perp (1 + \epsilon_L) \, , \quad \vert \epsilon_L \vert \ll 1 \, ,
\ee
where $\epsilon_\kappa$ and $\epsilon_L$ are two variables of small amplitude. Then, at first order we have
\be 
f(L_\perp) = f(k_\perp) + \epsilon_L k_\perp {\partial f(k_\perp) \over \partial k_\perp} \, , 
\ee
and also at first order 
\begin{eqnarray}
T(k_\perp,k_\perp (1 + \epsilon_L),k_\perp (1 + \epsilon_\kappa)) &=& 
{S \over B_0} E^{-s}_\perp(k_\perp) k_\perp \epsilon_L 
\left[k_\perp {\partial E^s_\perp(k_\perp) \over \partial k_\perp} - E^s_\perp(k_\perp)\right]  \nonumber \\
&=& {S \over B_0} E^{-s}_\perp(k_\perp) k_\perp^3 \epsilon_L 
{\partial \over \partial k_\perp} (E^s_\perp(k_\perp) / k_\perp) \, .
\end{eqnarray}
Therefore, equation (\ref{E3}) may be reduced at first order as 
\begin{eqnarray}
&& \partial_t \int_0^{+\infty} E^s_\perp(k_\perp) f(k_\perp) dk_\perp \nonumber \\
&=&- {1 \over B_0} \int_0^{+\infty} \int_{-\epsilon}^{+\epsilon} \int_{-\epsilon}^{+\epsilon} 
\left[ {S \over 2} E^{-s}_\perp(k_\perp) k_\perp^6 \epsilon_L^2 
{\partial \over \partial k_\perp} (E^s_\perp(k_\perp) / k_\perp) \right] {\partial f(k_\perp) \over \partial k_\perp} \, 
dk_\perp d\epsilon_\kappa d\epsilon_L \nonumber \\
&=&- {1 \over B_0} 
\int_0^{+\infty} \left[ C_\perp  E^{-s}_\perp(k_\perp) k_\perp^6 
{\partial \over \partial k_\perp} (E^s_\perp(k_\perp) / k_\perp)\right] {\partial f(k_\perp) \over \partial k_\perp} \, 
dk_\perp  \, ,
\label{E4}
\end{eqnarray}
where 
\be
C_\perp \equiv \int_{-\epsilon}^{+\epsilon}  \int_{-\epsilon}^{+\epsilon}  
{S \over 2} \epsilon_L^2  d\epsilon_\kappa d\epsilon_L \, ,
\label{Cp}
\ee
with $\epsilon$ a small positive number. The introduction of $\epsilon$ is necessary to ensure the strong 
locality of the interactions and therefore the convergence of integral (\ref{Cp}). An integration by parts gives
\begin{eqnarray}
\partial_t \int_0^{+\infty} E^s_\perp(k_\perp) f(k_\perp) dk_\perp 
= {1 \over B_0} \int_0^{+\infty} {\partial \over \partial k_\perp} \left[ C_\perp  E^{-s}_\perp(k_\perp) k_\perp^6 
{\partial \over \partial k_\perp} (E^s_\perp(k_\perp) / k_\perp)\right] f(k_\perp) \, 
dk_\perp  \, .
\label{E5}
\end{eqnarray}
Note that this operation implies for the function $f$ some constrains of convergence at the boundaries. 
Since $f$ is an arbitrary function one can write 
\begin{eqnarray}
\partial_t E^s_\perp(k_\perp) = {C_\perp \over B_0} 
{\partial \over \partial k_\perp} \left( k_\perp^6 E^{-s}_\perp(k_\perp) 
{\partial \over \partial k_\perp} \left({E^s_\perp(k_\perp) \over k_\perp} \right) \right) \, ,
\label{E7}
\end{eqnarray}
which is the differential limit of the wave turbulence equation for shear-Alfv\'en waves. It is useful to get an 
evaluation of $C_\perp$ since the constant in front of an equation always enters into account in the evaluation 
of the time scale dynamics. By noting that for strongly local interactions the angles of the triads are approximately 
$\pi/3$, one obtains
\be
C_\perp = {\pi \sqrt{3} \over 16} \int_{-\epsilon}^{+\epsilon}  \int_{-\epsilon}^{+\epsilon}  
\epsilon_L^2  d\epsilon_\kappa d\epsilon_L = {\pi  \epsilon^4 \over 4 \sqrt{3}} \, .
\label{e31}
\ee
Clearly, the degree of locality will strongly modify the time scale. For example, for strictly local interactions
$\epsilon=0$ and no evolution of the spectra is expected. This is consistent with the original equation 
(\ref{shearEq}) for which the right hand side is trivially zero if $L_\perp = k_\perp$. 

The same type of analysis for pseudo-Alfv\'en waves gives 
\begin{eqnarray}
\partial_t E^s_\parallel(k_\perp) = {C_\parallel \over B_0} 
{\partial \over \partial k_\perp} \left( k_\perp^6 E^{-s}_\perp(k_\perp) 
{\partial \over \partial k_\perp} \left({E^s_\parallel(k_\perp) \over k_\perp} \right) \right) \, ,
\label{E8}
\end{eqnarray}
where $C_\parallel \equiv \int_{-\epsilon}^{+\epsilon} \int_{-\epsilon}^{+\epsilon} (\pi/2) 
\epsilon_L^2 \sin \phi \, d\epsilon_\kappa d\epsilon_L = \pi  \epsilon^4 / \sqrt{3}$.

\section{Finite flux solutions and Komogorov constants}
\label{s5}

Equations (\ref{E7}) and (\ref{E8}) are the main results of the paper. They describe respectively the 
dynamical evolution of perpendicular and parallel fluctuations to the background magnetic field ${\bf B_0}$. 
We see that in the differential limit of strongly local interactions the wave turbulence equations are 
much simpler. They still satisfy the finite flux solutions as we will see below by looking at the power law 
solutions for shear-Alfv\'en waves. 
First of all let us introduce the energy flux $P^s_\perp(k_\perp)$ which is by definition
\be 
\partial_t E^s_\perp(k_\perp) \equiv - {\partial P^s_\perp(k_\perp) \over \partial k_\perp} \, .
\label{flux1}
\ee
We obtain
\be 
P^s_\perp (k_\perp) = - {C_\perp \over B_0}  k_\perp^6 E^{-s}_\perp(k_\perp) {\partial \over \partial k_\perp} 
\left({E^s_\perp(k_\perp) \over k_\perp} \right) \, .
\label{flux1b}
\ee
We shall find the power law solutions by introducing $E^s_\perp = C^s {k_\perp}^{n_s}$ into (\ref{flux1}); 
one gets
\be 
P^s_\perp (k_\perp) = - {C_\perp \over B_0} C^s C^{-s} (n_s -1) k_\perp^{n_s+n_{-s}+4} \, .
\label{flux1}
\ee
Therefore, the finite flux solutions, $P^s_\perp (k_\perp) = P^s_\perp = \rm{constant}$, correspond to 
\be
n_+ + n_- = - 4  \, , 
\label{co1}
\ee
with by symmetry 
\be
C^s C^{-s} = {P^s_\perp B_0 \over C_\perp (1- n_s)} = {P^{-s}_\perp B_0 \over C_\perp (1- n_{-s})}
= {B_0 \over C_\perp} \sqrt{{P^s_\perp P^{-s}_\perp \over (1- n_s)(1- n_{-s})} } \, ,
\ee
which leads to 
\begin{eqnarray}
E^+_\perp (k_\perp) E^-_\perp (k_\perp) &=& {B_0 \over C_\perp} \sqrt{{1 \over (1-n_+)(1-n_-)}} 
\sqrt{P^+_\perp P^-_\perp} k_\perp^{-4} \nonumber \\
&=& {B_0 \over C_\perp} \sqrt{{1 \over 5 + n_+ n_-}} 
\sqrt{P^+_\perp P^-_\perp} k_\perp^{-4} \nonumber \\
&=& {4 \sqrt{3} B_0 \over \pi \epsilon^4} \sqrt{{1 \over 5 + n_+ n_-}} 
\sqrt{P^+_\perp P^-_\perp} k_\perp^{-4} \, .
\end{eqnarray}
For balance turbulence, $n_+=n_-=-2$, and the finite flux solution is
\begin{eqnarray}
E_\perp (k_\perp) &=& \sqrt{B_0 \over 3 C_\perp} \sqrt{P_\perp} k_\perp^{-2} \nonumber \\
&=& \sqrt{4 B_0 \over \pi \sqrt{3} \epsilon^4} \sqrt{P_\perp} k_\perp^{-2} \nonumber \\
&\simeq& 0. 857 {\sqrt{B_0} \over \epsilon^2} \sqrt{P_\perp} k_\perp^{-2} \, .
\end{eqnarray}
We see that the Kolmogorov constant depends on an arbitrary truncation of the integration domain in equation 
(\ref{e31}). We remind that the Kolmogorov constant found by \citet{Galtier2000} for balance turbulence was, 
$C_K=0.585 \sqrt{B_0}$; therefore the constants coincide for $\epsilon \simeq 1.21$. Although the previous 
value violates the assumption of smallness for $\epsilon$ it could be taken {\it a posteriori} to find a unique 
solution compatible with the exact derivation of \citet{Galtier2000}.

A similar analysis for pseudo-Alfv\'en waves (\ref{E8}) gives the energy flux 
\be 
P^s_\parallel (k_\perp) = - {C_\parallel \over B_0}  k_\perp^6 E^{-s}_\perp(k_\perp) {\partial \over \partial k_\perp} 
\left({E^s_\parallel(k_\perp) \over k_\perp} \right) \, .
\label{flux2}
\ee
By introducing $E^s_\parallel = {\tilde C}^s {k_\perp}^{m_s}$ we find the finite flux solutions $m_s+n_{-s}=-4$
and thus 
\be
m_+ + m_- = - 4 \, , 
\label{co2}
\ee
\be
{\tilde C}^+ {\tilde C}^- = 
{B_0 C_\perp \over C_\parallel^2} {\sqrt{5+n_+n_-} \over 5+m_+m_-}
{P^+_\parallel P^-_\parallel \over \sqrt{P^+_\perp P^-_\perp}} \, , 
\ee
and 
\be
E^+_\parallel (k_\perp) E^-_\parallel (k_\perp) = {B_0 C_\perp \over C_\parallel^2} 
{\sqrt{5+n_+n_-} \over 5+m_+m_-} {P^+_\parallel P^-_\parallel \over \sqrt{P^+_\perp P^-_\perp}} 
k_\perp^{-4} \, ,
\ee
which reduces for balance turbulence to 
\begin{eqnarray}
E_\parallel (k_\perp) &=& \sqrt{B_0 C_\perp \over 3 C_\parallel^2} {P_\parallel \over \sqrt{P_\perp}} 
k_\perp^{-2} \nonumber \\
&=& \sqrt{B_0 \over 4 \pi \sqrt{3} \epsilon^4} {P_\parallel \over \sqrt{P_\perp}} k_\perp^{-2} \nonumber \\
&\simeq& 0.214 {\sqrt{B_0} \over \epsilon^2} {P_\parallel \over \sqrt{P_\perp}} k_\perp^{-2} \, .
\end{eqnarray}
Note that the Kolmogorov constant found by \citet{Galtier2000} for balance turbulence was, 
$C'_K=0.0675 \sqrt{B_0}$; in this case the constants coincide for $\epsilon \simeq 1.78$. 

Additionally, equations (\ref{E7}) and (\ref{E8}) reproduce the thermodynamic equilibrium solutions which 
correspond to zero flux \citep{Galtier2000}. In this case, it is straightforward to show from (\ref{E7}) and (\ref{E8}) 
that 
\be
n_s = m_s = 1 \, . 
\ee
As explained above, equations (\ref{E7}) and (\ref{E8}) are different from the model equations (73)--(74) 
derived in \cite{LG03} (where the notation are different; a comparison is possible if one takes 
$ke^{\uparrow \downarrow} \sim E_\perp^{+-}$) which do not give the thermodynamic equilibrium solutions.
The new system systematically derived here improved therefore the previous description while keeping the 
simplicity of a diffusion model. 

The nonlinear diffusion equations (\ref{E7})--(\ref{E8}) for non-zero cross-helicity may exhibit different power laws 
as finite flux solutions. Our knowledge of the initial system (\ref{shearEq})--(\ref{pseuEq}) imposes a priori that 
the power law indices satisfy the condition $-3<n_s , m_s<-1$ \citep{Galtier2000}. However, if we look at the 
diffusion equations we do not find any other constrain than relations (\ref{co1}) and (\ref{co2}) which means 
that in principle the power law indices are not bounded. The simplicity of the diffusion equations allows us to 
write a simple relation for shear-Alfv\'en energy fluxes, namely
\be
{P_\perp^+ \over P_\perp^-} =  {E^-_\perp \partial (E^+_\perp / k_\perp) / \partial k_\perp \over
E^+_\perp \partial (E^-_\perp / k_\perp) / \partial k_\perp} = 
{\partial \ln(E^+_\perp / k_\perp) / \partial k_\perp \over \partial \ln(E^-_\perp / k_\perp) / \partial k_\perp} \, .
\ee
In the stationary state, we obtain
\be
{P_\perp^+ \over P_\perp^-} = {n_+ - 1 \over n_- - 1}  = - \left({n_+ - 1 \over n_+ +5}\right) \, ,
\ee
which gives $P_\perp^+ / P_\perp^- = 1/2$ for $n_+=-1$ and $P_\perp^+ / P_\perp^- = 2$ for $n_+=-3$.
Note that the zero cross-helicity case corresponds to $n_+=n_-=-2$ for which $P_\perp^+=P_\perp^-$. 
In the general case which includes nonlocal interactions, we remind that we have $P_\perp^+ / P_\perp^- = 0$ 
for $n_+=-1$ and $P_\perp^+ / P_\perp^- = + \infty$ for $n_+=-3$ \citep{Galtier2000}. 
Therefore, the differences found between both predictions (from the diffusion and the integro-differential equations) 
give an evaluation of the nonlocal contributions.

\section{Numerical illustrations}
\label{num}

In order to check if the constant flux solutions are attractive we have performed two numerical simulations 
of the nonlinear diffusion equations. Only the case of shear-Alfv\'en waves (transverse fluctuations) has been 
considered. A linear viscous term is added in order to introduce a sink for the energy. In practice, the following 
equations are simulated 
\be
\partial_t E^{\pm}_\perp(k_\perp) = 
{\partial \over \partial k_\perp} \left( k_\perp^6 E^{\mp}_\perp(k_\perp) 
{\partial \over \partial k_\perp} \left({E^{\pm}_\perp(k_\perp) \over k_\perp} \right) \right) 
- \nu k_\perp^2 E^{\pm}_\perp(k_\perp) \, ,
\label{simu}
\ee
where $\nu$ is the viscosity (a unit magnetic Prandtl number is chosen). This type of equations is favorable 
to the use of a logarithmic subdivision of the $k_\perp$ axis such that in our case 
\begin{equation}
{k_\perp}_i = 2^{i/10} \ , 
\end{equation}
where $i$ is a positive integer. Such a discretization allows us to reach Reynolds numbers much greater than 
in direct numerical simulations. We take $i_{max}=200$ which corresponds to a ratio of about $10^6$ 
between the largest and the smallest scales. In our simulations the viscosity is fixed to $\nu=5 \times 10^{-5}$. 

\begin{figure}
\plotone{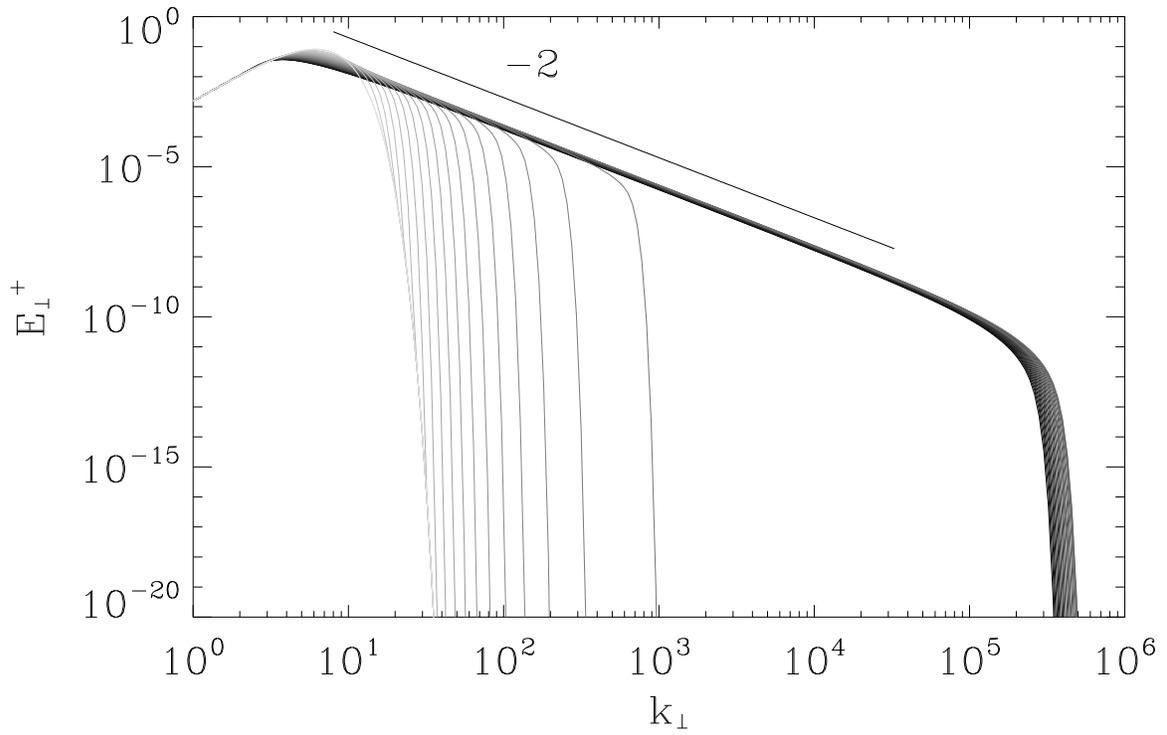}
\caption{Time evolution of the energy spectrum $E^+_\perp(k_\perp)$ for a cross-helicity $\rho=0$. 
The stationary spectrum is well fitted by a power law in $k_\perp^{-2}$. Forty spectra are shown with a constant 
interval of time.}
\label{Fig2}
\end{figure}
\begin{figure}
\plotone{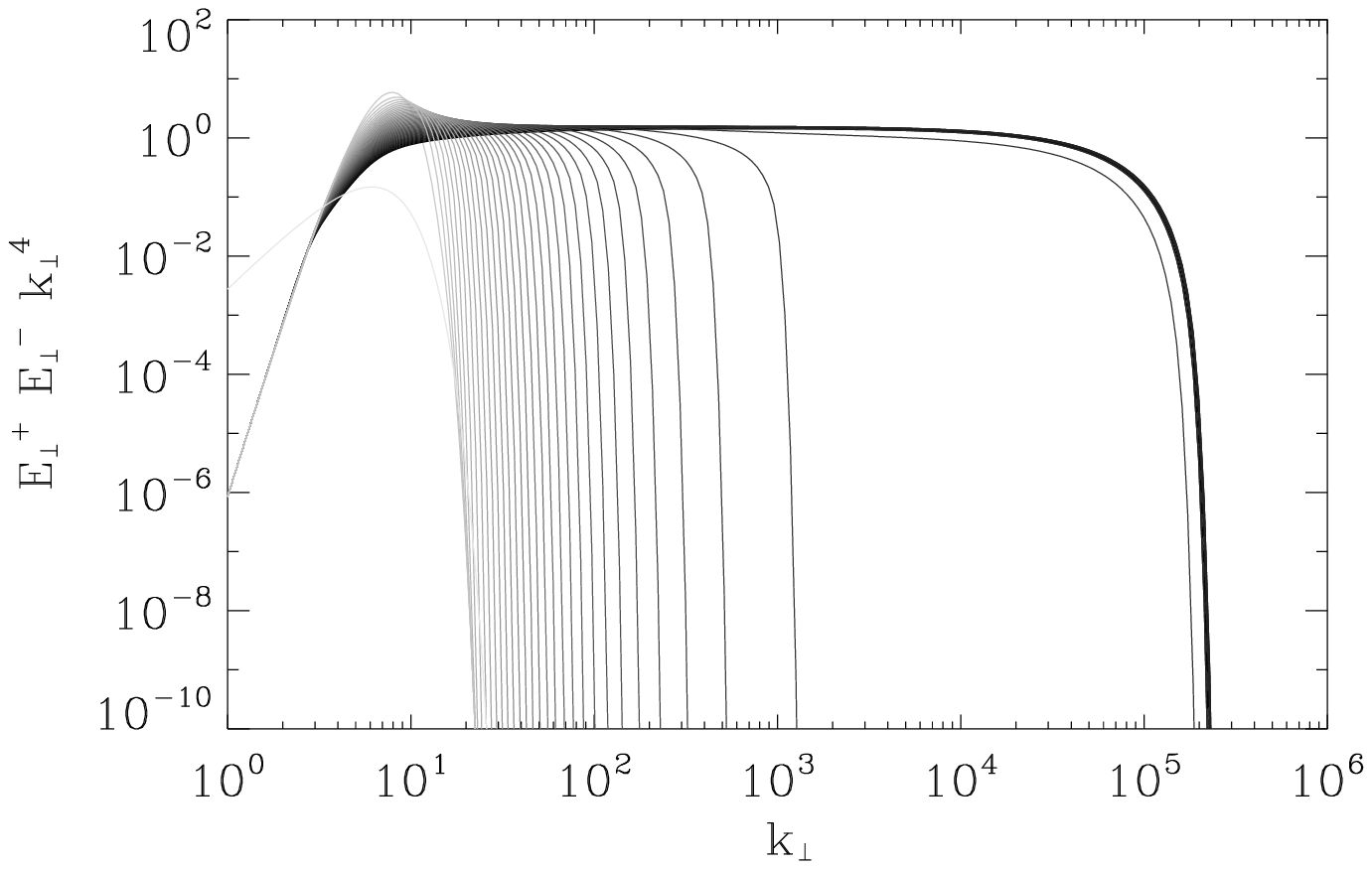}
\caption{Time evolution of the compensated energy spectra $E^+_\perp E^-_\perp \kpe^4$ for a cross-helicity 
$\rho=0.8$. The stationary spectra satisfy the relation $n_+ + n_- = -4$. Forty spectra are shown with a constant 
interval of time.}
\label{Fig3}
\end{figure}

The first simulation corresponds to the zero cross-helicity case for which by definition $E^+_\perp=E^-_\perp$.
Large scale spectra centered around $k_\perp = k_0$ are taken initially with the form
\be
E_\perp^\pm(\kpe) \sim \kpe^3 \exp(-\kpe^2/k_0^2) \, ,
\label{CI}
\ee
where $k_0=5$. Only the time evolution of $E_\perp^+$ is given in Fig. \ref{Fig2}. We see that the front of 
the energy spectrum propagates towards larger wavenumbers to reach eventually a $k_\perp^{-2}$--stationary 
spectrum as predicted by the theory. We  may note the acceleration of the front until the dissipation scale is 
reached since spectra are separated by a constant interval of time. 
In the second simulation we have fixed initially the (reduced) cross-helicity to
\be
\rho = {E_\perp^+ - E_\perp^- \over E_\perp^+ + E_\perp^-} = 0.8 \, . 
\label{rho1}
\ee
Like in the previous case the initial spectra are centered around $k_\perp = k_0=5$ with more energy in spectrum 
$E_\perp^+$ than in $E_\perp^-$. We keep the same form as (\ref{CI}). The result is shown in Fig. \ref{Fig3}. 
We see that the compensated spectra fit well with the theoretical prediction $k_\perp^{-4}$ over several decades. 
In fact, spectra meet at relatively small $k_\perp$ and exhibit the same inertial range with the same 
$\kpe^{-2}$--spectrum over a wide range of scales.

\section{Discussion}
\label{s6}

It would be relevant to investigate if whether or nor the system recently derived by \citet{Bill09} for strong 
(anisotropic) MHD turbulence is able to recover 
the present equations when the limit of wave turbulence is taken. It would also be interesting to analyze if an 
anomalous scaling is detected during the front propagation (which is not easy to find here). We remind that the 
wave kinetic equations (\ref{shearEq}) exhibit a $\kpe^{-7/3}$--spectrum during the non-stationary phase (at 
zero cross-helicity) which is still not understood \citep{Galtier2000}. Anomalous scalings are weaker in diffusion 
models of turbulence than in wave kinetic equations \citep{colm10}. We plan to further investigate this point 
by using for example a higher order numerical scheme. We also plan to further compare numerically the 
nonlinear diffusion equations and the wave kinetic equations to determine the domain of divergence between 
them or the influence of an external force \citep[see \eg][]{GN08}. 

The nonlinear diffusion equations for non-zero cross-helicity (\ref{E7})--(\ref{E8}) is a simple and therefore 
useful system for describing a wide class of astrophysical plasmas. The solar corona with the myriad of 
magnetic loops which are characterized by a strong axial magnetic field is probably a good example of  
application of Alfv\'en wave turbulence \citep{Rappazzo,Bigot08b}. This regime is also relevant for coronal 
holes where the solar wind is produced. For both examples, equations (\ref{E7})--(\ref{E8}) could give a 
description of MHD turbulence at large scales since it seems inevitable that at smaller scales the strong 
turbulence regime overcomes. In this case a coupling with for example the advection-diffusion model proposed 
by \cite{chandran08} would be relevant. 

\acknowledgments

We acknowledge Institut universitaire de France for financial support.



\begin{thebibliography}{}
\bibitem[Alexakis et al.(2007)]{Alexakis}
Alexakis, A., Bigot, B., Politano, H., \& Galtier, S. 2007, Phys. Rev. E, 76, 056313
\bibitem[Bigot et al.(2008a)]{Bigot08a}
Bigot, B., Galtier, S., \& Politano, H. 2008a, Phys. Rev. E, 78, 066301
\bibitem[Bigot et al.(2008b)]{Bigot08b}
Bigot, B., Galtier, S., \& Politano, H. 2008b, \aap, 490, 325
\bibitem[Bigot et al.(2008c)]{Bigot08c}
Bigot, B., Galtier, S., \& Politano, H. 2008c, \prl, 100, 074502
\bibitem[Buchlin and Velli(2007)]{buchlin}
Buchlin, \'E., and Velli, M. 2007, \apj, 662, 701
\bibitem[Chae et al.(1998)]{chae}
Chae, J., Sch\"uhle, U., \& Lemaire, P. 1998, \apj, 505, 957
\bibitem[Chandran(2005)]{chandran}
Chandran, B.D.G. 2005, \prl, 95, 265004
\bibitem[Chandran(2008)]{chandran08}
Chandran, B.D.G. 2008, \apj, 685, 646
\bibitem[Connaughton and Nazarenko(2004)]{colm}
Connaughton, C., \& Nazarenko, S. 2004, \prl, 92, 044501
\bibitem[Connaughton and Newell(2010)]{colm10}
Connaughton, C., \& Newell, A.C. 2010, \pre, 81, 036303
\bibitem[Cranmer and van Ballegooijen(2003)]{Cranmer03}
Cranmer, S.R., \& van Ballegooijen, A.A. 2003, \apj, 594, 573
\bibitem[Cranmer(2010)]{Cranmer10}
Cranmer, S.R. 2010, \apj, 710, 676
\bibitem[David et al.(1998)]{david}
David, C., Gabriel, A.H., Bely-Dubau, F., Fludra, A., Lemaire, P., \& Wilhelm, K. 1998, A\&A, 336, L90
\bibitem[Davidson(1958)]{davidson}
Davidson, B. 1958, Neutron transport theory, Oxford University Press, Oxford, 1958
\bibitem[Dyachenko et al.(1992)]{Dyachenko}
Dyachenko, S., Newell, A.C., Pushkarev, A.N., \& Zakharov, V.E. 1992, Physica D, 57, 96
\bibitem[Galtier(1999)]{Galtier1999}
Galtier, S. 1999, \apj, 521, 483
\bibitem[Galtier et al.(2000)]{Galtier2000}
Galtier, S., Nazarenko, S.V., Newell, A.C., \& Pouquet, A. 2000, J. Plasma Phys., 63, 447
\bibitem[Galtier et al.(2002)]{Galtier2002}
Galtier, S., Nazarenko, S.V., Newell, A.C., \& Pouquet, A. 2002, \apjl, 564, L49
\bibitem[Galtier(2006)]{Galtier06a}
Galtier, S. 2006, J. Plasma Phys., 72, 721
\bibitem[Galtier and Chandran(2006)]{Galtier06b}
Galtier, S., \& Chandran, B.D.G. 2006, Phys. Plasmas, 13, 114505
\bibitem[Galtier(2008)]{Galtier2008}
Galtier, S. 2008, Phys. Rev. E, 77, 015302
\bibitem[Galtier(2009)]{Galtier2009a}
Galtier, S. 2009, Nonlin. Processes Geophys., 16, 83
\bibitem[Galtier(2009)]{Galtier2009b}
Galtier, S. 2009, \apj, 704, 1371
\bibitem[Galtier and Nazarenko(2008)]{GN08}
Galtier, S., \& Nazarenko, S.V. 2008, J. Turbulence, 9(40), 1
\bibitem[Heisenberg(1948)]{Hen}
Heisenberg, W. 1948, Proc. R. Soc. Lond. A, 195, 402
\bibitem[Heyvaerts and Priest(1992)]{HP92}
Heyvaerts, J., \& Priest, E.R. 1992, \apj, 390, 297
\bibitem[Iroshnikov(1964)]{Iro}
Iroshnikov, P. 1964, Sov. Astron., 7, 566
\bibitem[Kolmogorov(1941)]{K41}
Kolmogorov, A.N. 1941, Dokl. Akad. Nauk SSSR, 32, 16
\bibitem[Kraichnan(1963)]{kraichnan63}
Kraichnan, R.H. 1963, Phys. Fluids, 6, 1603
\bibitem[Kraichnan(1965)]{Krai}
Kraichnan, R.H. 1965, Phys. Fluids, 8, 1385
\bibitem[Leith(1967)]{leith}
Leith, C.E. 1967, Phys. Fluids, 10, 1409
\bibitem[Lithwick and Goldreich(2003)]{LG03}
Lithwick, Y., \& Goldreich, P. 2003, \apj, 582, 1220
\bibitem[MacBride et al.(2008)]{macbride}
MacBride, B.T., Smith, C.W., \& Forman, M.A. 2008, \apj, 679, 1644
\bibitem[Matthaeus et al.(1996)]{Bill96}
Matthaeus, W.H., Ghosh, S., Oughton, S., \& Roberts, D.A. 1996, J. Geophys. Res., 101, 7619
\bibitem[Matthaeus et al.(2009)]{Bill09}
Matthaeus, W.H., Oughton, S., \& Zhou, Y. 2009, Phys. Rev. E, 79, 035401
\bibitem[Mininni and Pouquet(2007)]{mininni}
Mininni, P.D., \& Pouquet, A. 2007, \prl, 99, 254502
\bibitem[Orszag and Kruskal(1968)]{orszag}
Orzsag, S.A., \& Kruskal, M.D. 1968, Phys. Fluids, 11, 43
\bibitem[Podesta(2009)]{podesta}
Podesta, J.J. 2009, \apj, 698, 986
\bibitem[Pouquet et al.(1976)]{pouquet}
Pouquet, A., Frisch, U., \& L\'eorat, J. 1976, J. Fluid. Mech., 77, 321
\bibitem[Rappazzo et al.(2007)]{Rappazzo}
Rappazzo, A.F., Velli, M., Einaudi, G., \& Dahlburg, R.B. 2007, \apj, 657, L47
\bibitem[Sahraoui et al.(2009)]{Sahraoui}
Sahraoui, F., Goldstein, M.L., Robert, P., \& Khotyaintsev, Yu.V. 2009, \prl, 102, 231102
\bibitem[Salem et al.(2009)]{salem}
Salem, C., Mangeney, A., Bale, S.D., \& Veltri, P. 2009, \apj, 702, 537
\bibitem[Scalo and Elmegreen(2004)]{elmegreen}
Scalo, J., \& Elmegreen, B.G. 2004, \araa, 42, 275
\bibitem[Smith et al.(2006)]{Smith06}
Smith, C.W., Hamilton, K., Vasquez, B.J., \& Leamon, R.J. 2006, \apj, 645, L85
\bibitem[Zhou and Matthaeus(1990)]{zhou}
Zhou, Y., \& Matthaeus, W.H. 1990, J. Geophys. Res., 95, 14881
\end{thebibliography}
\end{document}